\documentclass[12pt]{article}
\usepackage{epsfig}

\date{\today}

\begin{document}
\begin{center}
{\Large\bf Inflationary 
solutions with a five dimensional \\complex scalar field }
\\
\vspace{0.5cm}
{\bf J.J.\ van der Bij$^{1}$ } and~ 
{\bf Eugen Radu$^{2}$ } 
\\
\vspace*{0.2cm}
{\it $^{1}$Albert-Ludwigs-Universit\"at Freiburg, Institut f\"ur 
Physik, 
\\Hermann-Herder-Stra\ss e 3, D-79104 Freiburg, Germany}
\\
{\it $^2$ Department of
Mathematical Physics, National University of Ireland Maynooth, 
Ireland}
\vspace{0.5cm}
\end{center}
\begin{abstract}
We discuss inflationary solutions of the coupled 
Einstein-Klein-Gordon equations 
for a complex field in a five dimensional spacetime 
with a compact $x^5$-dimension.
As a new feature, the scalar field contains a dependence on 
the extra dimension of the
form $\exp(i m x^5) $, corresponding to
Kaluza-Klein excited modes.
In a four dimensional picture, a nonzero $m$ implies the presence of 
a new term in the scalar field potential.
An interesting feature of these solutions is the possible existence 
of several periods of oscillation of the scalar field
around the equilibrium value at the minimum of the potential.
These oscillations lead to cosmological periods of accelerated 
expansion  of the universe.
\end{abstract}
%
\section{Introduction}
The inflationary paradigm is currently seen as a solution
to the problems arising in standard hot big-bang cosmology 
\cite{Guth:1980zm,Linde},
such as flatness, horizon and 
structure formation.
A necessary condition for inflation to occur is that the energy
density of the early stage of the universe be dominated by the vacuum 
energy of a 
quasi-homogeneous scalar field, usually called the "inflaton".

After the original inflationary scenario was proposed,
various models have been investigated to obtain a natural inflation.
A possibility is to consider higher dimensional gravity 
theories of the Kaluza-Klein type
coupled to a scalar field 
(see e.g. \cite{Holman:1990hg}-\cite{Mazumdar:1999tk}).
After reduction to four dimensions, theories originally formulated 
in $4+D$ dimensions take a Jordan-Brans-Dicke form,
and generically contain also scalar fields of geometrical origin, 
describing
the size of extra dimensions.
The usual Einstein gravity can be obtained
using a suitable conformal transformation. 
 
Most of the studies on this subject assume that the inflaton field 
represents
the symmetries of the extra- $D-$dimensional compact manifold.
However, the situation may be more complicated, as we point out in 
this letter.
Considering for simplicity only one extra-dimension,
we suppose that the complex inflaton field has a dependence on the 
$x^5$-th dimension of the form $\exp (imx^5)$. 
The compactness of the extra-coordinate forces the constant $m$ 
to take certain discrete values.
However, all the observable quantities are $x^5$-independent, since 
the 
action of the generator of $x^5$  translation is equivalent 
to the action of a rigid phase transformation. 

Also, it is well know that the presence of scalar fields in cosmology 
introduces 
the problem of the appropriate potential,
a problem which has been studied extensively in the literature.
We find that a $x^5-$dependence of the inflaton field will induce a 
new 
scalar potential term
 in the four dimensional Lagrangian,
which may influence behavior of the solutions.
By numerically solving the field
equations  we find that, for a double-well inflaton potential,
there  is a succession of short bursts of inflation,
after a generic initial period of inflation.
A nonzero $m$ generally enhances this effect.

As a new feature, we speculate about the possibility that $m$ changes 
its value 
for solutions with scalar field nodes.
This leads to new qualitative features, the most interesting being
the generic existence of multiple
periods of accelerated expansion of the universe, for a transition to 
a higher $m$.
 
\section{General framework}
\subsection{Action and equations of motion}

We consider the following action principle in $4+1$ dimensions
\begin{eqnarray}
\label{S5}
S_5=\int d^5x \sqrt{-g_{(5)}}\Big( 
\frac{1}{2 \kappa^2} R_5+L_m\Big),
\end{eqnarray}
where $g_{\mu \nu}$ and $\kappa^2$ are the five-dimensional metric  
and 
gravitational constant.
$L_m$ is the Lagrangian for a complex scalar field $\Psi$, with a 
potential $V$ depending 
on $|\Psi|^2=\Psi^*\Psi$ only
\begin{eqnarray}
\label{Ls}
L_m= - g^{\mu \nu}\partial_{\mu}\Psi^*\partial_{\nu}\Psi- V (\Psi) .
\end{eqnarray}
Variation of the action with respect to $g^{\mu \nu}$  yields the 
Einstein equations
\begin{eqnarray}
\label{E-eqs}
R_{\mu \nu}-\frac{1}{2}R g_{\mu \nu}=\kappa^2~T_{\mu \nu},
\end{eqnarray}
where the energy momentum tensor is given by
\begin{eqnarray}
\label{tik}
T_{\mu \nu}=\Psi_{,\mu}^{\star}\Psi_{,\nu}+
\Psi_{,\nu}^{\star}\Psi_{,\mu}-g_{\mu \nu}
\left( \Psi_{,\rho}\Psi^{,\rho}+V(\Psi)\right).
\end{eqnarray}
The scalar field equation is \begin{eqnarray}
\label{KG-eqs}
\Big(\nabla^2-\frac{d V}{d |\Psi|^2}\Big)\Psi=0.
\end{eqnarray}
\subsection{The ansatz}
We take a five-dimensional metric ansatz with no dependence on the 
extra-dimension $x^5$, 
on the form
\begin{eqnarray}
\label{metric}
ds_5^2=-dt^2+a^2(t)d x^i d x^i+\Phi^2(t)(d x^5)^2,
\end{eqnarray}
where $a(t)$ is the scale factor of the three-dimensional space, 
$\Phi(t)$ is the scale
factor of the extra-dimension and $x^i$ denotes the three space 
directions.
Here we suppose a compact fifth dimension, with $0\leq \chi < L$.
 The corresponding nonvanishing components of the Einstein tensor 
$G_{\mu \nu}=R_{\mu \nu}-\frac{1}{2}R g_{\mu \nu}$ read
\begin{eqnarray}
\label{Et}
G_i^i=-\frac{\dot{a}^2+2a\ddot{a}}{a^2}-\frac{2\dot{a}\dot{\Phi}+a\ddot{\Phi}}{a\Phi},
~~
G_t^t=-\frac{3}{a^2\Phi}(\Phi\dot{a}^2+a\dot{a}\dot{\Phi}),~~
G_5^5=-\frac{3}{a^2 }( \dot{a}^2+a\ddot{a}),
\end{eqnarray}
where a dot denotes derivative with respect to time.

The usual scalar field ansatz considered in the literature to obtain 
solutions 
compatible with the symmetries of the  line element (\ref{metric}) is 
a homogeneous one, with
$\Psi=\chi(t)$.
However, a dependence of the $\Psi$-field on the extra-dimension may 
be introduced by 
taking
\begin{eqnarray}
\label{psi}
\Psi=\frac{1}{\sqrt{2}}\chi(t) e^{i m x^5}
\end{eqnarray}
where $\chi(t)$ is a real function, this matter ansatz being still 
compatible 
with the symmetries of the metric (\ref{metric}).
Since scalar fields must be single-valued functions with respect to 
$x^5$, we find
$\Psi(t, x^5)=\Psi(t, x^5+L)$. From this 
periodicity, values of $m$ must be of the form
$m=2\pi n/L$, where $n$ is an integer,
representing the winding number with 
respect to the 
extra-coordinate $x^5$.
While $x^5/L$ covers the trigonometric circle once, 
the field winds $n$ times around.

  It appears natural to consider a generalization of the 
simple matter ansatz (\ref{psi}) consisting
in a superposition of $N$ individual modes
\begin{eqnarray}
\label{psigen}
\Psi=\frac{1}{\sqrt{2}}\sum_{k=0}^{N-1}\chi_k(t) e^{i m_k x^5},
\end{eqnarray}
with $\chi_k(t)$ real functions and $m_k=2\pi k/L$.
However, one can prove that 
this general ansatz is not compatible with the assumption that the
five-dimensional line element (\ref{metric}) does not depend on 
$x^5$.
A straightforward computation gives 
\begin{eqnarray}
\label{rels}
\nonumber
\Psi\Psi^{\star}=\frac{1}{2}\big(\sum_{k=1}^{N-1}\chi_k^2(t) +
\sum_{\scriptstyle k,j=0\atop\scriptstyle k\ne j}^{N-1}
\cos((m_k-m_j)x^5)\chi_k (t)\chi_j (t)\big),
\\
\Psi_{,t}\Psi_{,t}^{\star}=\frac{1}{2}\big(\sum_{k=1}^{N-1}\dot{\chi}_k^2(t) +
\sum_{\scriptstyle k,j=0\atop\scriptstyle k\ne j}^{N-1}
\cos((m_k-m_j)x^5)\dot{\chi}_k (t)\dot{\chi}_j (t)\big),
\\
\nonumber
\Psi_{,x^5}\Psi_{,x^5}^{\star}=\frac{1}{2}\big(\sum_{k=1}^{N-1}m_k^2\chi_k^2(t) +
\sum_{\scriptstyle k,j=0\atop\scriptstyle k\ne j}^{N-1}
\cos((m_k-m_j)x^5)m_km_j\chi_k (t)\chi_j (t)\big),
\end{eqnarray}
implying from (\ref{tik}) a dependence of the energy momentum tensor 
on the 
extra-coordinate, which is not compatible with the Einstein tensor 
(\ref{Et}). 

Therefore, rather surprising, only one single mode can be excited 
within the metric ansatz (\ref{metric}). Thus, for the rest of this 
paper, we consider
a scalar field $\Psi=\frac{1}{\sqrt{2}}\chi(t) \exp(i m x^5)$, with 
$m=2\pi n/L $ 
and an arbitrary integer $n$. 

With these conventions, the nonvanishing components of the 
energy-momentum tensor are
\begin{eqnarray}
\label{Tik}
\nonumber
T_i^i&=&p_4=\frac{1}{2}\dot{\chi}^2-m^2 
\frac{\chi^2}{2\Phi^2}-V(\chi),
\\
-T_t^t&=&\rho=\frac{1}{2}\dot{\chi}^2+m^2 
\frac{\chi^2}{2\Phi^2}+V(\chi),
\\
\nonumber
T_5^5&=&p_5=\frac{1}{2}\dot{\chi}^2+m^2 
\frac{\chi^2}{2\Phi^2}-V(\chi).
\end{eqnarray}
One can see that a nontrivial dependence on the extra dimension
increases the energy density $\rho$ and pressure component associated 
with 
the extra-dimension $p_5$, while decreasing the four dimensional 
pressure $p_4=T_i^i$.

The dynamical equations for $a,\Phi$ and $\chi$ (with $H=a'/a$) are
\begin{eqnarray}
\label{eqs}
\nonumber
\frac{\ddot{\Phi}}{\Phi}+3H\frac{\dot{\Phi}}{\Phi}&=&
\frac{\kappa^2}{3}\big(7m^2\frac{\chi^2}{\Phi^2}+4V(\chi)\big),
\\
H^2+H\frac{\dot{\Phi}}{\Phi}&=&\frac{\kappa^2}{3}
\big(\dot{\chi}^2+m^2\frac{\chi^2}{\Phi^2}+2V(\chi)\big),
\\
\nonumber
\ddot{\chi}+(3H+\frac{\dot{\Phi}}{\Phi})\dot{\chi}
+m^2 \frac{\chi}{\Phi^2}+
\frac{\partial V(\chi)}{\partial \chi}&=&0.
\end{eqnarray}
By using suitable combinations of the Einstein equations, 
we find the following useful relations
\begin{eqnarray}
\label{aux-rel}
\frac{\ddot{a}}{a}&=&H\frac{\dot{\Phi}}{\Phi}-
\frac{\kappa^2}{3}\left(\dot{\chi}^2+ 
m^2\frac{\chi^2}{\Phi^2}\right),
~~~~
\ddot{a}=-\frac{\kappa^2}{3}2a^2 p_5,
\\
\nonumber
\frac{d^2 }{dt^2}(a^3  \Phi) &=&\frac{\kappa^2}{3}\Phi a^3 
\big(3m^2 \frac{\chi^2}{ \Phi^2}+8V(\chi)\big),
~~~
\frac{d }{dt}(a \Phi)=\frac{\kappa^2}{3} a \Phi \rho,
~~~
\frac{d }{dt}(\frac{a}{\Phi})=\frac{\kappa^2}{3}\frac{a}{\Phi}
(\dot{\chi}^2+m^2\frac{\chi^2}{\Phi^2}),
\end{eqnarray}
which can be used to predict general features of the solutions.
For example, one can prove that the scale factor $a$ of the 
three-dimensional
space is  always increasing,  $\dot{a}>0$.
 Also, for periods of accelerated expansion $(\ddot{a}>0)$, the 
extra-dimension radius  
$\Phi(t)$ is a strictly increasing function $\dot{\Phi}>0$. 

\subsection{Four dimensional reduction}
With this ansatz, it is possible to reduce (\ref{S5}) to a four 
dimensional Lagrangian.
The resulting Lagrangian mimics the Lagrangian of a 
Jordan-Brans-Dicke theory where
there is an extra scalar field (related to the size of the 
extra-dimension),
nonminimally coupled to the four dimensional Ricci scalar. 

The resulting action after the dimensional reduction 
for the metric ansatz (\ref{metric}) is
\begin{eqnarray}
\label{S4-e}
S_4=\int d^4x \sqrt{-g_4}
\Big( 
\frac{1}{2 \kappa^2} \Phi R_4-\frac{\Phi}{2}( 
\chi_{,\mu}\chi^{,\mu}+m^2\frac{\chi^2}{\Phi^2}+2V(\chi))
\Big).
\end{eqnarray}
One can see that a nonzero winding number implies a direct coupling 
between extra-dimension
radius and the scalar $\chi$, while 
the resulting field equations are still given by (\ref{eqs}). 

The above system  admits also an alternative picture in an "Einstein 
frame", with
a minimal coupling between the extra-dimension radius and 
four dimensional gravity.
These are complementary pictures and the solution in one frame can be 
directly translated 
to the results in the other frame by field and time coordinate 
redefinitions.

If we note $\Phi=\exp{(\alpha \phi)}$ (with 
$\alpha=\sqrt{2\kappa^2/3})$ and consider a 
 five-dimensional metric of the form \cite{Overduin:1998pn}
\begin{eqnarray}
\label{d5}
ds_5^2=e^{-\alpha\phi}ds_4^2+e^{2\alpha\phi}d \chi^2
\end{eqnarray}
the reduced four dimensional Lagrangian in the Einstein frame reads 
\begin{eqnarray}
\label{S4}
S_4=\int d^4x \sqrt{-g_4} \Big( 
\frac{R_4}{2\kappa^2} -\frac{1}{2}\partial_{\mu}\phi 
\partial_{\mu}\phi
-\frac{1}{2}\partial_{\mu}\chi \partial_{\mu}\chi-U(\phi,\chi)
\Big),
\end{eqnarray}
where 
\begin{eqnarray}
\label{U}
U(\phi,\chi)=e^{-3\alpha\phi}m^2\chi^2+e^{- \alpha\phi}V(\chi).
\end{eqnarray}
The four dimensional metric is taken to be: 
\begin{eqnarray}
\label{m4}
ds_4^2=-d\tau^2+\bar{a}^2(\tau)d \vec{x} d \vec{x}
\end{eqnarray}
which implies the following relations between the five- and 
four-dimensional functions:
\begin{eqnarray}
\label{rel}
\tau=\int dt ~\Phi(t),~~~~~~~~~~\bar{a}=\frac{a(t)}{\Phi(t)}.
\end{eqnarray}
The field equations in  Einstein  frame and a line element (\ref{m4}) 
read
\begin{eqnarray}
\nonumber
\bar{H}^2&=&\frac{\kappa^2}{2}(\phi'^2+\chi'^2-U),
\\
\chi''+3\bar{H}\chi'&=&-\frac{\partial U}{\partial \chi}
=-e^{\alpha \phi} \frac{\partial V}{\partial \chi}-e^{-3 \alpha \phi} 
m^2 \chi,
\\
\nonumber
\phi''+3\bar{H}\phi'&=&-\frac{\partial U}{\partial \phi}
=\alpha(e^{\alpha \phi} V + \frac{3}{2}e^{-3 \alpha \phi} m^2 
\chi^2),
\end{eqnarray}
where a prime denotes derivative with respect to $\tau$ and 
$\bar{H}=\bar{a'}/\bar{a}$.
If we suppose that asymptotically $\chi\to \chi_0$ such that 
$V(\chi_0)=0$, 
while  $\phi$ takes a finite value,
the term $e^{-3\alpha\phi}m^2\chi^2$ will correspond to an 
effective four-dimensional cosmological constant.

\section{Inflationary solutions}
\subsection{Slow roll approximation}
Inflationary solutions of the equations (\ref{eqs})
have been discussed for $m=0$  by some authors in the context of 
inflation from higher dimensional theories 
(see e.g. \cite{Holman:1990hg}-\cite{Mazumdar:1999tk}).

It can be seen immediately that an exact solution 
exists for $\chi_0$ satisfying the condition
\begin{eqnarray}
\left(m^2\frac{\chi }{\Phi^2}+
\frac{\partial V}{\partial \chi}\right)\Big|_{\chi=\chi_0}=0.
\end{eqnarray}
Therefore, if $V(\chi_0) \neq 0$,
the scale factor
$a(t)$ expands exponentially, similar to the radius of the 
extra-dimensions,
$a(t) \sim e^{ct}$, with $c^2=\kappa^2V(\chi_0)/6 $.
Also, it can easily be proven from (\ref{aux-rel})
that there are no solutions with a constant extra-dimension radius.

The crucial ingredient of nearly all known inflationary scenarios is 
a period of
"slow roll" evolution of the inflaton field.
During this period $\chi$ changes very slowly, so that its kinetic 
energy $\dot{\chi}^2/2$
remains always  much smaller than its potential energy.

For the situation discussed in this paper, the slow-roll 
approximation reads
\begin{eqnarray}
\ddot{\chi} \ll  3H \dot {\chi},
~~~\ddot{\chi} \ll  \frac{\dot{\Phi}}{\Phi} \dot {\chi},~~~
\ddot{\Phi} \ll  3H \frac{\dot {\Phi}}{\Phi},
\end{eqnarray}
while the Kaluza-Klein modes are required to satisfy
\begin{eqnarray}
\label{cond1}
  m^2\frac{\chi }{\Phi^2}\ll 3H \dot {\chi},
  ~~ m^2\frac{\chi ^2}{\Phi^2}\ll V(\chi),
\end{eqnarray}
Note that these conditions imply a number of constraints 
on the parameters (for example
from (\ref{cond1}),  $m$ cannot take arbitrary large values). 

With these assumptions, one finds  the 
following approximate solution 
\begin{eqnarray}
\Phi=a^{2/3}+const.,~~H=\sqrt{\frac{\kappa^2}{3} V(\chi)},~~
t=-\frac{1}{3\kappa^2} f(\chi),
\end{eqnarray}
where 
\begin{eqnarray}
f(\chi)=\int~d \chi \frac{\sqrt{V(\chi)}}{V'(\chi)}.
\end{eqnarray}
The end of inflation is marked by the condition 
\begin{eqnarray}
\frac{1}{2}\dot{\chi}^2+m^2 \frac{\chi^2}{2\Phi^2}\simeq V(\chi).
\end{eqnarray}
\subsection{Numerical solutions}
However, the slow roll conditions are sufficient, but not necessary 
to maintain inflation.
There is the possibility that inflation  continues 
after the slow roll ends, during a period of fast oscillations
of the inflaton field (see e.g. \cite{Damour:1997cb}).
However, this implies a number of constrains on the inflaton 
potential.
Here we investigate the possibility that the supplementary term in 
the Lagrangian
induced by the inflaton field dependence on the extra dimension will 
lead to oscillations of the scalar fields, that translate to 
periods of accelerated expansion of the universe.

No analytic arguments are available in this case and 
the equations of motion (\ref{eqs}) should be solved numerically.
One may use the translation symmetry of the field equations
to set an arbitrary value for the initial time coordinate, which is 
taken $t_0=0$.
To obtain numerical solutions, one needs also to fix six initial 
conditions 
$(a(0),\dot{a}(0),\Phi(0),\dot{\Phi}(0),\chi(0),\dot{\chi}(0))$.
One of the Einstein equations is a constraint equation
that  allows one to express one of these constants
in terms of the others. 

The scalar field potential considered in this letter has 
the usual double-well form
\begin{eqnarray}
\label{pot1}
 V(\chi)= \lambda^2(\chi^2-v^2)^2.
\end{eqnarray}
Therefore, one should also specify the values of the constants 
appearing in the
(\ref{pot1}) as input parameters, as well as the value of $m$ 
(these are usually consider of order unity, while we take $L=2\pi$).

Following the usual approach and using a standard ordinary  
differential  
equation solver, we
evaluate  the  initial conditions at $t_0=10^{-14}$ and 
integrate  towards  $t\to\infty$, 
up to a maximal value $t_{max}$.

\subsubsection{Solutions with fixed winding number}
The case $m=0$ has been the subject of many studies. A detailed 
study of the solutions of the Einstein equations
with a real scalar field has been done in particular by Belinski et 
al. 
\cite{Belinsky:1985zd} (see also \cite{Scialom:1994uq}).

Here we do not aim at finding precise quantitative features of
 the model presented in Section 2.
Instead we are looking  for qualitatively new features introduced by 
a nonzero value of $m$.
Thus the field equations (\ref{eqs})  have been solved 
for several values of $(m,\lambda,v)$ and a large set of initial 
conditions, looking for generic
 properties (for example, the solutions plotted here have 
$\lambda=1$).

For all values of the parameters, the scale factor $a(t)$ is a 
strictly increasing function
(as proven analitically).
After an initial period of faster expansion, the scale function
$a(t)$ becomes  proportional to 
the function $\sqrt{t}$.
We could not find 
configurations with a constant value of $a(t)$ for large enough 
values of $t$.

The scalar field behavior  is also generic.
Independent on the initial conditions, the scalar field always 
approaches asymptotically the
vacuum expectation value $v$,
after a number of oscillations  around $v$ (see Figure 1a). 
 
The behavior of the extra-dimension radius $\Phi(t)$ is somewhat 
special.
Here, for some set of initial conditions we notice the existence of a 
local maximum
 of the radius of the 
\newpage
\setlength{\unitlength}{1cm}

\begin{picture}(8,8)
\centering
\put(2,0){\epsfig{file=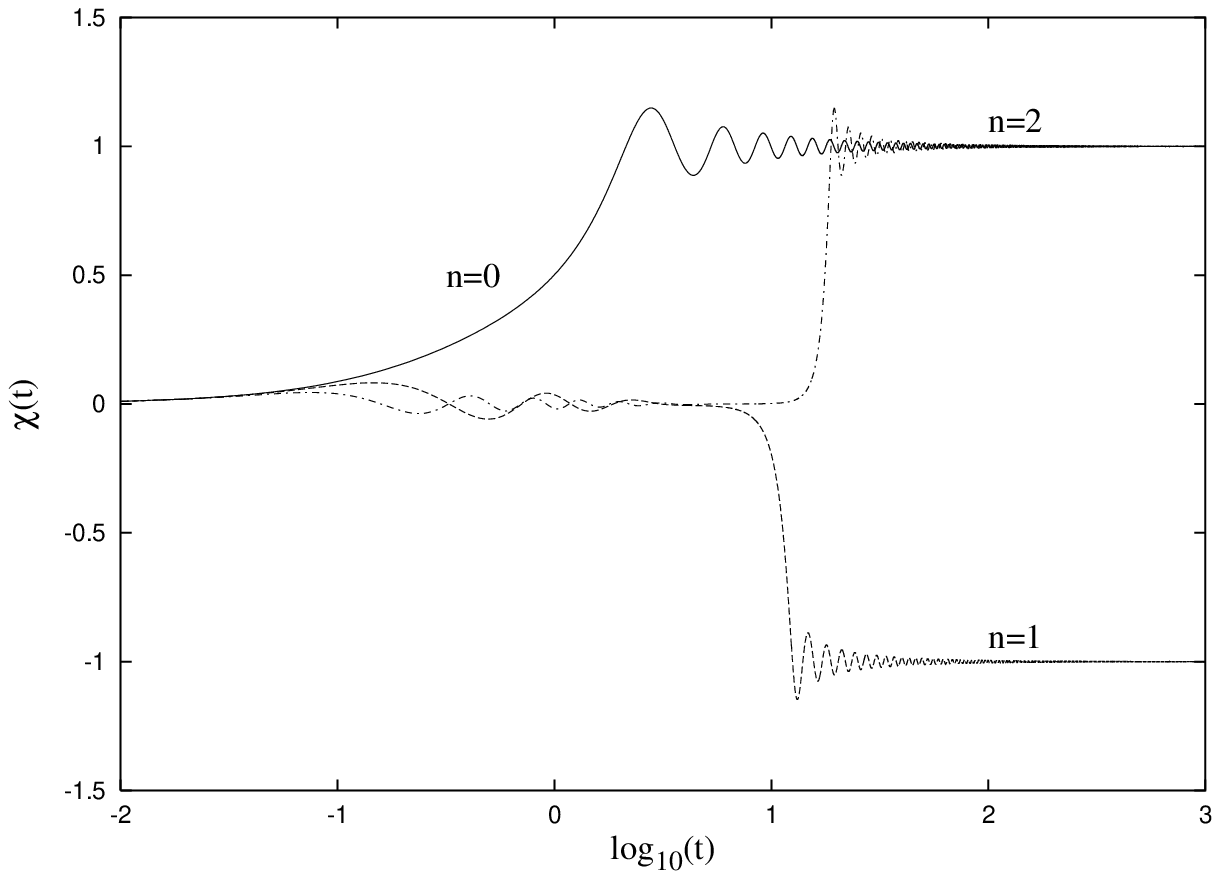,width=12cm}}
\end{picture}
\begin{picture}(19,8.5)
\centering
\put(2.7,0){\epsfig{file=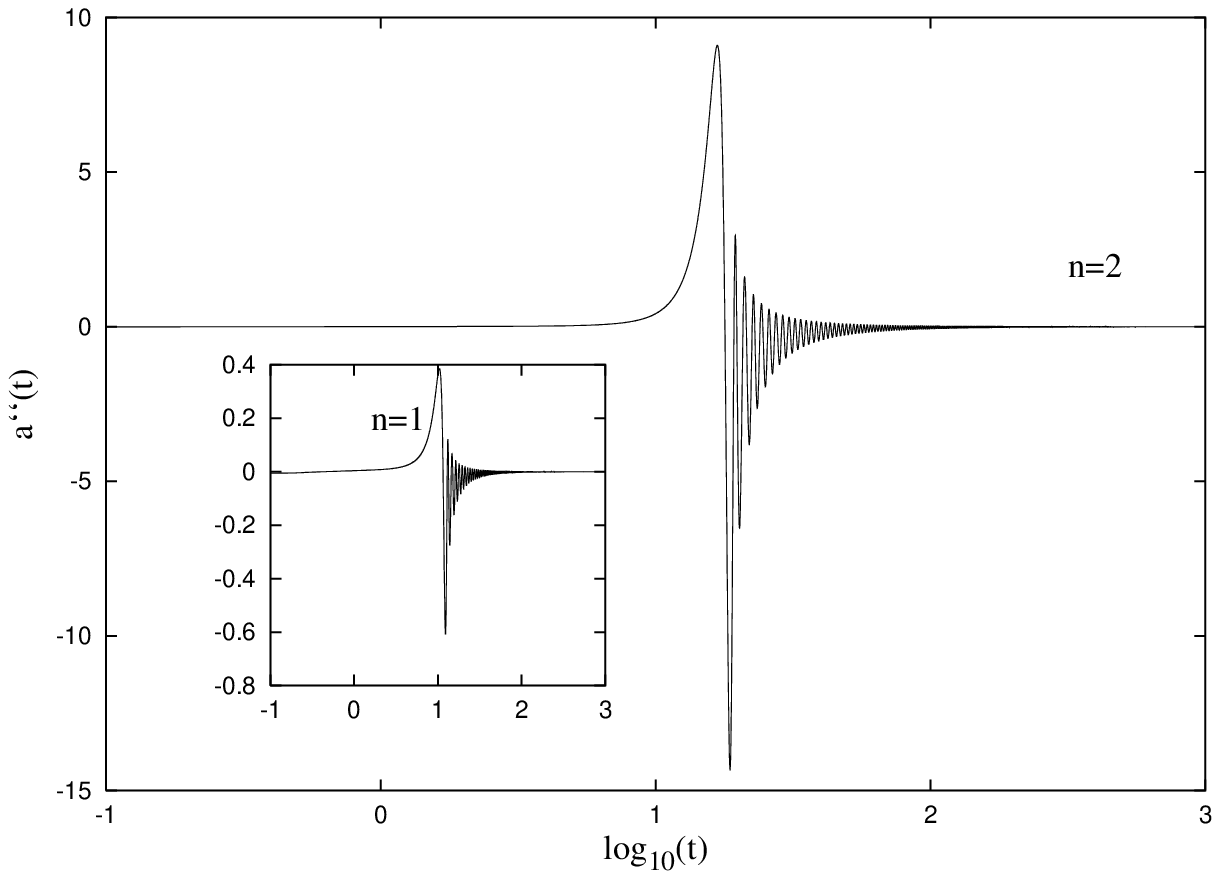,width=12cm}}
\end{picture}
\\
{\small {\bf Figure 1.}
The scalar field $\chi$ and the second derivative of the scale factor 
$a$  
 are represented  as a function of $t$ for typical solutions.
}
\\
\\
extra-dimension, 
and a more complicated behavior of this function for small values of 
$t$
(the function $\chi(t)$ should be a strictly increasing function in 
this case).
However, for large enough values of $t$, $\Phi(t)$ is a strictly 
increasing function,
$\Phi(t) \sim t$.
We could not find solutions  where the extra-dimension radius 
approaches asymptotically
a constant value.

The second derivative of the function $a(t)$ 
(which is an indication for the existence of inflation) has
 an interesting behavior for $m\neq 0$.
For most of the considered configurations,
the oscillations of the scalar translate to a number of oscillations 
around zero of the function 
\newpage
\setlength{\unitlength}{1cm}

\begin{picture}(18,8)
\centering
\put(1,0.0){\epsfig{file=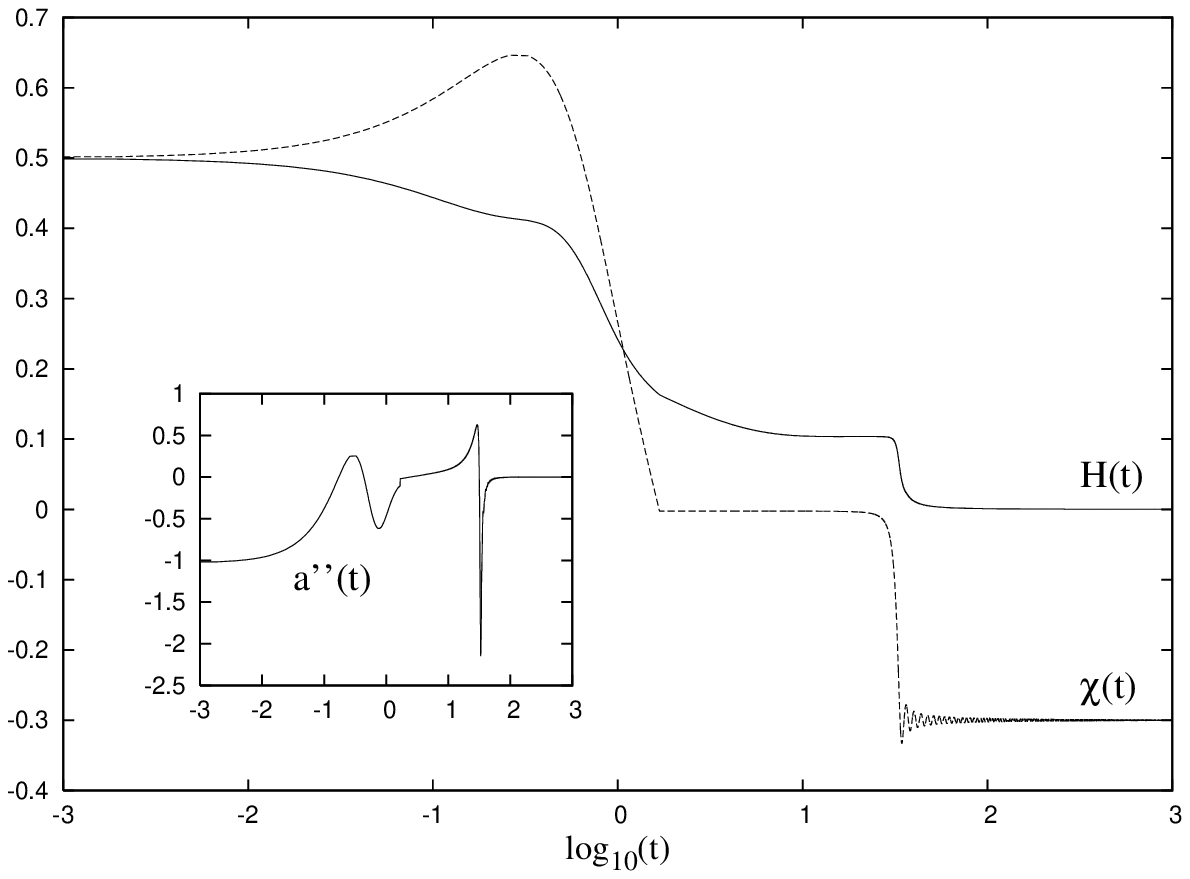,width=12cm}}
\end{picture}
\\
\\
{\small {\bf Figure 2.}
The second derivative of the metric function $a$, the scalar field 
$\chi$ 
and the Hubble parameter $H=a^{-1}da/dt$
 are represented  as a function of $t$ for a typical solution 
 with winding number transition from $m_i=0$ to $m_f=3$.
 }
\\
\\
$\ddot{a}$ (see Figure 1b).
This implies therefore the existence of several periods of inflation
separated by periods of noninflationary
expansion.
This "multiple inflation" behavior is a consequence of 
the form of  the inflaton field potential.
Although this effect is present also
for $m=0$, its details
 depend on the value of $m$.
For example, the length and magnitude of $\ddot{a}$ oscillations 
generally strongly increase with $m$.
This effect is maximized for initial value of the scalar field near 
the
false vacuum $\chi=0$.
Also, $\ddot{a}$ vanishes always for large values of $t$,
while the expansion rate $H$ is a decreasing function.

\subsubsection{Solutions with  winding number transitions}
 In the previous section we took $m$ to be  a constant.
If one takes $\lambda$ to be very large this is a very natural 
assumption,
since in the limit $\lambda \to \infty$, $m$ becomes a 
topologically conserved charge. 
  However, we remark that 
for the physical situation that we consider here with  finite $\lambda$, 
 there is no reason why the value of $m$  
should be a conserved quantity. There is no charge associated with 
it. 

 It appears interesting to take $m$ as a fluctuating variable
and to look for possible physical effects.
A change of $m$ for an arbitrary $t$ is unacceptable, since it 
generally leads to 
discontinuities of the metric functions, 
which are considered unphysical \cite{Hawking}.
This can be avoided by considering solutions with nodes, 
where the scalar field $\chi$ crosses the
axes for some values of the time coordinate $t=t_1,t_2,..$.
Note that this requires a choice of the  parameters in the problem, 
since nodeless
solutions are present as well.

We solve the field equations for a number of configurations, 
starting with a configuration with winding number $m_i$ and
changing the value of $m$
from $m_i$ to $m_f$
as the scalar field $\chi$ crosses the axes for some $t$.
Therefore, this does not lead to discontinuities of the relevant 
functions $(a, \chi, \Phi)$.
 
The results we have found for $m_f<m_i$ are rather similar to the 
case of a fixed winding number. 
The  behavior of solutions following the changing of the winding 
number
is very similar to the 
  solutions with a fixed $m=m_i$, with small quantitative  
differences only.
New qualitative features are found for transitions to a higher 
winding number.
We find that  generically
this implies the existence of a second period of inflation, whose 
length
increases with $m_f$ (see Figure 2).
We can understand intuitively this effect  by remarking that
a transition of the scalar field to a higher (lower) winding number 
corresponds 
to injecting (extracting)  extra energy in the system,
as can be seen from the field equations.

The typical inflaton field behavior is  presented in Figure 2
as well as the evolution of the expansion rate $H$ 
(which is a decreasing function).
After changing the winding number, the scalar field
 presents very small oscillations around 
zero, followed by a sudden transition towards $v$
at some $t>t_1$, approaching asymptotically the vacuum expectation 
value.
The metric functions $a,\Phi$ are strictly increasing quantities,
apparently not noticing the changing of $m$.

These results are generally not sensitive to the initial conditions,
but are affected by the scalar field potential parameters $\lambda, 
v$.
However, we found that a similar behavior appears generically 
for a large enough difference $m_f-m_i$.

One can also imagine more complicated scenarios, with 
several $m$-transitions, leading to a succession of bursts of 
inflation.

\section{Conclusions}
The purpose of this work was to consider 
the possibility that the bulk inflaton field
in a higher dimensional theory possesses  a 
dependence on the extra-dimensions.
In the simplest five-dimensional case, this introduces a 
new parameter in the theory, corresponding to a winding number $m$ 
with respect the
$x^5$-direction. 

After reduction to four dimensions, the dependence on the 
extra-dimension
is apparent as  a new term in the inflaton potential. 
We have presented numerical arguments that, for a nonzero winding 
number,
the solutions of the Einstein-Klein-Gordon equations present a number 
of new qualitative features.
The possibility that   $m$ changes the value 
for solutions with scalar field nodes has been also considered.

Therefore, the inclusion of a inflaton 
dependence on the extra-dimensions leads to a rich model
with new curious features which deserve further 
investigation.
For example, it would be interesting to study the quantitative
details as well as density fluctuations within this model. 
The scenario can easily be generalized for a number $D>1$ 
of extra-dimensions with a torus compactification.
\newpage
{\bf Acknowledgement}
\\
The work of ER was supported by the Graduiertenkolleg of the Deutsche
Forschungsgemeinschaft (DFG): Nichtlineare Differentialgleichungen;
Modellierung, Theorie, Numerik, Visualisierung and by the 
Enterprise--Ireland Basic
Science Research Project SC/2003/390 of Enterprise-Ireland.

\end{document}